\documentclass[12pt]{article}
\usepackage{graphicx}
\usepackage{amsfonts}
\usepackage{amsmath,amssymb}
\usepackage{mathrsfs}
\usepackage{setspace}
\usepackage{cite}

\def\beq{\begin{equation}}
\def\eeq{\end{equation}}
\def\bea{\begin{eqnarray}}
\def\eea{\end{eqnarray}}
\def\noi{\noindent}

\def\ba{\begin{array}}
\def\ea{\end{array}}

\begin{document}

\begin{center}
{\Large \bf \sf 
Clusters of bound particles in a quantum integrable many-body system 
and number theory}

\vspace{1.3cm}

{\sf B. Basu-Mallick$^1$\footnote{e-mail address:
bireswar.basumallick@saha.ac.in},
Tanaya Bhattacharyya$^2$\footnote{e-mail address:
tanaya.bhattacharyya@googlemail.com}
and Diptiman Sen$^3$\footnote{e-mail address: diptiman@cts.iisc.ernet.in}}
\bigskip

{\em $^1$Theory Division, Saha Institute of Nuclear Physics, \\
1/AF Bidhan Nagar, Kolkata 700 064, India}

\bigskip

{\em $^2$Department of Physics, St. Xavier's College, \\
30 Park Street, Kolkata 700 016, India}

\bigskip

{\em $^3$Centre for High Energy Physics, Indian Institute of Science, \\
Bangalore 560 012, India}
\end{center}

\bigskip
\bigskip

\vskip 1.1 cm

\noi {\bf Abstract}
\medskip

We construct clusters of bound particles for a quantum integrable derivative
$\delta$-function Bose gas in one dimension. It is found that clusters of bound
particles can be constructed for this Bose gas for some special values of
the coupling constant, by taking the quasi-momenta associated with the 
corresponding Bethe state to be equidistant points on a {\it single} circle 
in the complex momentum plane. Interestingly, there exists a connection 
between the above mentioned special values of the coupling constant and some 
fractions belonging to the Farey sequences in number theory. 
This connection leads to a classification
of the clusters of bound particles for the derivative $\delta$-function 
Bose gas and the determination of various properties of these clusters like
their size and their stability under a variation of the coupling constant.

\newpage
\section{Introduction}
\renewcommand{\theequation}{1.{\arabic{equation}}}
\setcounter{equation}{0}
\medskip
 
One-dimensional (1D) quantum integrable many-body systems
with short range interactions have emerged as an active area of research
\cite{LL63}-\cite{CCGOR11},
due to their effectiveness
in describing recent experiments using strongly
interacting ultracold atomic gases 
\cite{K02}-\cite{KWW06}. 
Indeed, many results of these
experiments have been understood within the framework of the 
1D quantum integrable
Lieb-Liniger model or the $\delta$-function Bose gas, which can be solved 
exactly through the Bethe ansatz.

In this context it may be recalled that, for a large class of quantum 
integrable systems, the method of coordinate Bethe ansatz directly yields exact 
eigenfunctions in the coordinate representation. One can study the 
asymptotic form of these eigenfunctions in the limit of infinite length of
the system (i.e., when all $x_i$'s are allowed to take value
in the range $-\infty < x_i < \infty $). 
If the probability density associated with an eigenfunction decays sufficiently
fast when the relative distance between any two particle coordinates tends 
towards infinity (for a translationally invariant system),
a bound state is formed. It is well known,
for the case of the $\delta$-function Bose gas with $N \geq 2$,
that bound states 
exist for all negative values
of the coupling constant \cite{Mc64,Ya68,Th81,Fa80,Sk90,KBI93,Ta99}.
The quasi-momenta associated with such a bound state are represented by 
equidistant points lying on a straight line or `string' parallel to 
the imaginary axis in the complex momentum plane. 
For the case of the $\delta$-function Bose gas with negative values of
the coupling constant, one can also construct Bethe eigenfunctions 
corresponding to more complex structures like clusters of bound particles. 
The quasi-momenta corresponding to such clusters of bound particles are 
represented through discrete points lying on several `strings', 
all of which are parallel to the imaginary axis in the complex momentum 
plane \cite{Ya68,Ta99,CC07,CC07a,Do10,PS11}.

Similar to the case of the $\delta$-function Bose gas mentioned above,
there exists another exactly solvable and quantum integrable bosonic system
with a Hamiltonian given by 
\begin{equation} {\cal H}_N 
~=~ -\hbar^2 ~\sum_{j=1}^N ~\frac{\partial^2}{\partial x_j^2} ~+~ 2i
\hbar^2 \eta ~\sum_{l<m} ~\delta(x_l - x_m )~ \Big( \frac{\partial}{\partial
x_l} + \frac{\partial} {\partial x_m} \Big) \, , 
\label{a2} \end{equation}
where we have chosen $2m=1$ and 
$\eta$ is a real (nonzero) dimensionless coupling constant 
for this choice of mass
\cite{Gu87,KB93,SMB94,BB02,BB03}. The Hamiltonian (\ref{a2}) of this
derivative $\delta$-function Bose gas can be obtained by projecting that of
an integrable derivative nonlinear Schr\"{o}dinger (DNLS) quantum
field model on the $N$-particle subspace. 
Classical and quantum versions of such DNLS field
models have found applications in different areas of physics like circularly
polarized nonlinear Alfven waves in plasma, quantum properties of optical
solitons in fibers, and in some chiral Tomonaga-Luttinger liquids obtained
from the Chern-Simons model defined in two dimensions 
\cite{CLL79,KN78,MP96,WSKI78,Cl92,RS91,AGJPS96}. 
The scattering and bound states of the derivative $\delta$-function
Bose gas (\ref{a2}) have been studied extensively by
using the methods of coordinate as well as algebraic Bethe ansatz
\cite{Gu87,KB93,SMB94,BB02,BB03,BBS03,BBS04,BBS04a}. 
It turns out that for the cases $N=2$ and $N=3$, bound
states of this model can be constructed for any value of $\eta$ within 
its full range: 
$0 < \mid \eta \mid < \infty$. However, for any given value of $N \geq 4$,
the derivative $\delta$-function Bose gas allows bound 
states in only certain non-overlapping ranges of the coupling constant
$\eta$ (the union of these ranges yields a proper subset of the full range of
$\eta$),
and such non-overlapping ranges of $\eta$
can be determined by using the Farey sequences in number theory
\cite{BBS03,BBS04,BBS04a}. 

In analogy with the case of the $\delta$-function Bose gas, one may 
think that clusters of bound particles can only be constructed
for the case of derivative $\delta$-function Bose gas by properly assigning 
the corresponding quasi-momenta on several concentric
circles or circular `strings' in the complex momentum plane.
However, we have recently found that, for the Hamiltonian 
(\ref{a2}) with some special values of the coupling constant $\eta$, 
clusters of bound particles can be constructed in a much simpler way 
by assigning the corresponding quasi-momenta as equidistant points on a
{\it single} circle in the complex momentum 
plane \cite{BBS13}. The purpose of the present article is mainly to review 
the key results of Ref.~\cite{BBS13} and make some additional comment.
The arrangement of this article is as follows. 
In Sec.~2, we first discuss the general form of Bethe eigenstates
for the derivative $\delta$-function Bose gas. Then 
we identify a sufficient condition for which such a Bethe eigenstate
would represent clusters of bound particles. In Sec.~3, 
we classify all possible solutions of this sufficient 
condition and obtain different types of clusters of bound particles
for the derivative $\delta$-function Bose gas.
In Sec.~4 we discuss various properties of such clusters of bound particles, 
such as the sizes of the clusters, their stability under the variation of the 
coupling constant. We end with some concluding 
remarks in Sec.~5. 

\section{Construction of clusters of bound particles}
\renewcommand{\theequation}{2.{\arabic{equation}}}
\setcounter{equation}{0}

In the coordinate representation, the eigenvalue equation for the
Hamiltonian (\ref{a2}) may be written as
\begin{eqnarray} 
{\cal H}_N ~\tau_N( x_1, x_2, \cdots , x_N ) ~=~ E ~\tau_N( x_1, x_2,
\cdots , x_N )~, \label{b1} \end{eqnarray}
where $\tau_N( x_1, x_2, \cdots , x_N )$ denotes a completely symmetric
$N$-particle wave function. Since ${\cal H}_N$ commutes with
the total momentum operator given by
\begin{equation} 
{\cal P}_N ~=~ -i\hbar ~\sum_{j=1}^N ~\frac{\partial}{\partial x_j} ~
\, ,
\label{b2} \end{equation}
$\tau_N( x_1, x_2, \cdots , x_N )$ can be chosen as a simultaneous
eigenfunction of these two commuting operators. 
Note that ${\cal H}_N$ remains 
invariant while ${\cal P}_N$ changes sign if we change the sign of $\eta$
and transform all the $x_i \rightarrow - x_i$ at the same time;
such a transformation may be called as `parity transformation'.
Due to the invariance of ${\cal H}_N$ under this parity transformation,
it is sufficient to study the eigenvalue problem (\ref{b1}) for
one particular sign of $\eta$, say, $\eta >0$. 
The eigenfunctions for $\eta < 0$ case can then be constructed from
those for $\eta>0$ case by simply changing $x_i \rightarrow -x_i$; 
this leaves all energy eigenvalues invariant but reverses the
sign of the corresponding momentum eigenvalues. 

For the purpose of solving the eigenvalue problem (\ref{b1})
through the coordinate Bethe ansatz, it is convenient to divide
the coordinate space $R^N \equiv \{ x_1, x_2, \cdots x_N \}$
into various $N$-dimensional sectors defined through inequalities like
$x_{\omega(1)}< x_{\omega(2)}< \cdots < x_{\omega(N)}$, where 
$\{\omega(1), \omega(2), $ $\cdots , \omega(N)\}$ 
represents a permutation of the integers $\{1,2, \cdots ,N \}$. Since the
interaction part of the Hamiltonian (\ref{a2}) vanishes within each such 
sector, the resulting eigenfunction can be expressed as a superposition of
free particle wave functions. The coefficients associated
with these free particle wave functions can be computed by using the
interaction part of the Hamiltonian (\ref{a2}), which is nontrivial only
at the boundary of two adjacent sectors.
In the region $x_1< x_2 < \cdots < x_N$, such eigenfunctions
can be written in the form \cite{Gu87,SMB94}
\begin{equation} \tau_N (x_1, x_2 , \cdots , x_N) ~=~
\frac{1}{\sqrt{N!}} \sum_\omega \left(\prod_{l<m}
\frac{A(k_{\omega(m)},k_{\omega(l)})}{A(k_m,k_l)}\right) \rho_{\omega(1),
\omega(2), \cdots , \omega(N)} (x_1, x_2, \cdots , x_N) ~, \label{b3} 
\end{equation}
where
\begin{equation} 
\rho_{\omega(1), \omega(2), \cdots , \omega(N)} (x_1, x_2, \cdots , x_N) ~=~
\exp ~\{ i (k_{\omega(1)}x_1 + \cdots + k_{\omega(N)} x_N ) \} ~,
\label{b4} \end{equation}
$k_n$'s are all distinct quasi-momenta, $\omega$ represents an element of
permutation group for the integers $\{ 1,2,....N \}$
and $\sum_{\omega}$ implies summing over all such permutations.
The coefficient $A(k_l,k_m)$ in Eq.~(\ref{b3}) is obtained 
by solving the two-particle problem related to the derivative $\delta$-function
Bose gas and this coefficient is given by
\begin{equation} 
A(k_l,k_m) ~=~ \frac{k_l - k_m + i \eta (k_l+ k_m)}{k_l - k_m} ~.
\label{b5} \end{equation}
The eigenvalues of
the momentum (\ref{b2}) and Hamiltonian (\ref{a2}) operators,
corresponding to the eigenfunctions $\tau_N(x_1, x_2, \cdots , x_N)$ of the
form (\ref{b3}), are easily obtained as 
\begin{eqnarray}
&&~~~~~~~~~~~~~~{\cal P}_N ~\tau_N(x_1, x_2, \cdots , x_N) ~=~ \hbar
\Big(\sum_{j=1}^N k_j \Big) ~\tau_N(x_1, x_2, \cdots , x_N) ~, \nonumber
~~~~~~~~~~~~~~~~~~~~~~ (2.6a) \\
&&~~~~~~~~~~~~~~{\cal H}_N ~\tau_N(x_1, x_2, \cdots , x_N) ~=~
\hbar^2 \Big(\sum_{j=1}^N k_j^2 \Big) ~\tau_N(x_1, x_2, \cdots , x_N) ~.
\nonumber ~~~~~~~~~~~~~~~~~~~~~ (2.6b) 
\end{eqnarray}
\addtocounter{equation}{1}
Next, we shall discuss how Bethe states of the form in (\ref{b3})
lead to the bound states of the derivative $\delta$-function Bose gas,
by allowing $k_j$'s to take complex values in an appropriate
way. As mentioned earlier, 
for a translationally invariant system, a wave function
represents a localized bound state if the corresponding
probability density decays sufficiently fast 
when any of the relative coordinates measuring the distance between
a pair of particles tends towards infinity.
To obtain the condition for which
the Bethe state (\ref{b3}) would represent such a localized bound state, 
let us first consider the following
wave function in the region $x_1<x_2<\cdots <x_N$ :
\begin{eqnarray} 
\rho_{1,2, \cdots ,N} (~ x_1,x_2, \cdots , x_N ~) ~=~ \exp
~(i\sum_{j=1}^N
k_j x_j)\,, \label{b7} \end{eqnarray}
where $k_j$'s in general are complex valued wave numbers.
Since the corresponding momentum eigenvalue 
given by $\hbar\sum_{j=1}^N k_j$ must be a real quantity, one
obtains the condition
\begin{equation} \sum_{j=1}^N q_j ~=~ 0 ~, \label{b8} \end{equation}
where $q_j$ denotes the imaginary part of $k_j$. By using (\ref{b8}), the 
probability density
for the wave function $\rho_{1,2, \cdots ,N} (~ x_1,x_2, \cdots
,x_N ~)$ in (\ref{b7}) can be expressed as
\begin{eqnarray} 
{|\rho_{1,2,\cdots ,N} (~ x_1,x_2, \cdots , x_N ~)|}^2 ~=~ \exp ~\Big\{ ~
2 \sum_{r=1}^{N-1} \Big(\sum_{j=1}^r q_j\Big) ~y_r ~\Big \} ~, \label{b9}
\end{eqnarray}
where the $y_r$'s are the $N-1$ relative coordinates: $y_r \equiv
x_{r+1} - x_r \, $. Hence, the
probability density in (\ref{b9}) decays exponentially in the limit
$y_r \rightarrow \infty$ for one or more values of $r$, provided that all
the following conditions are satisfied:
\begin{equation} 
q_1< 0 ~, ~~~~q_1+q_2 < 0 ~, ~~\cdots\cdots ~~, ~\sum_{j=1}^{N-1} ~q_j
< 0 ~. \label{b10} \end{equation}
It should be observed that the wave function (\ref{b7}) is obtained by taking
$\omega$ as the identity permutation in (\ref{b4}). However, the
Bethe state (\ref{b3}) also contains terms like (\ref{b4}) with 
$\omega$ representing all possible nontrivial permutations.
The conditions which ensure the decay of such a term, associated
with any nontrivial permutation $\omega$, are evidently given by
\begin{equation} q_{\omega(1)}< 0 ~, ~~~~q_{\omega(1)}+ q_{\omega(2)}< 0 ~, 
~~\cdots\cdots ~~, ~\sum_{j=1}^{N-1} ~q_{\omega(j)} < 0 ~. \label{b11} 
\end{equation}
It is easy to check that above conditions, in general, contradict the 
conditions given in Eq.~(\ref{b10}). To bypass this problem and ensure an
overall decaying wave function (\ref{b3}), it is sufficient 
to assume that the coefficients of all terms
$\rho_{\omega(1), \omega(2), \cdots , \omega(N)}
(x_1, x_2, \cdots , x_N)$ with nontrivial permutations
take the zero value. This leads to a set of relations given by
\begin{equation} A( k_{r}, k_{r+1} ) ~=~ 0 \, , \quad {\rm for} \quad r \in
{\Omega}_N\,,
\label{b12} \end{equation}
where ${\Omega}_N \equiv \{1,\,2,\,\cdots,\,N-1\}$.
Consequently, the simultaneous validity of the conditions
(\ref{b8}), (\ref{b10}) and (\ref{b12}) ensures that the Bethe state
$\tau_N(x_1, x_2, \cdots , x_N)$ (\ref{b3}) would represent a bound state.

Let us now analyse the conditions (\ref{b8}), (\ref{b10}) and
(\ref{b12}) for the 
case of the derivative $\delta$-function Bose gas.
Using the conditions (\ref{b8}) and (\ref{b12}) along with Eq.~(\ref{b5}),
one can easily derive an expression for all the quasi-momenta as
\begin{equation} k_n ~=~ \chi ~e^{-i(N+1-2n)\phi} ~, \label{b13} \end{equation}
where $\chi$ is a real, non-zero parameter, 
and $\phi$ is related to the coupling constant $\eta$ as
\begin{equation} 
\phi ~=~ \tan^{-1} (\eta ) ~ \Longrightarrow ~~ \eta = \tan \phi \, .
\label{b14} \end{equation}
To obtain an unique value of $\phi$ from the above equation,
it may be restricted 
to the fundamental region $-\frac{\pi}{2} < \phi (\neq 0) < \frac{\pi}{2}$.
Furthermore, since we have seen that ${\cal H}_N$ (\ref{a2}) remains invariant
under the `parity transformation', it is enough to study the corresponding
eigenvalue problem only within the range $0 < \phi <
\frac{\pi}{2}$. Next, we consider 
the remaining conditions (\ref{b10}) for the existence of a localized bound
state. 
Since summation over the imaginary parts of $k_n$'s in (\ref{b13}) yields
\begin{eqnarray} 
\sum_{j=1}^l q_j = -\chi ~\frac{\sin (l \phi)}{\sin \phi} ~\sin [(N-l)
\phi]\, , \label{b15} \end{eqnarray}
Eq.~(\ref{b10}) can be expressed as
\begin{equation} 
\chi ~\frac{\sin (l \phi)}{\sin \phi} ~\sin [(N-l) \phi] ~>~ 0 \, ,
\quad {\rm for} \quad l \in {\Omega}_N\, , \label{b16} \end{equation}
where ${\Omega}_N$ denotes the set of integers $\{1,\,2,\,\cdots,\,N-1\}$.
Consequently, for any given values of $\phi$ and $N$, a bound state
would exist when all the inequalities in Eq.~(\ref{b16}) 
are simultaneously satisfied for some real non-zero value of $\chi$.
In our earlier works it was shown that, for any given value of $N \geq 4$,
the derivative $\delta$-function Bose gas allows bound states in only certain 
non-overlapping ranges of the coupling constant $\phi$ called `bands'
and the location of these bands
can be determined exactly \cite{BBS03,BBS04a}.

We would now like to find the conditions for constructing clusters of bound
particles in the case of the derivative $\delta$-function Bose gas. To this
end,
we shall first discuss the concept of a `clustered state' for
any translationally invariant system, and then give a prescription
for finding the Bethe states representing clusters of bound particles.
Let us consider a system of $N$ particles which are divided
into some groups or clusters --- with at least one group containing more than
one particle. It is assumed that particles within the same group
behave like the constituents of a bound state, but particles corresponding
to different groups behave like the constituents of a scattering state. 
More precisely, a wave function corresponding to such an $N$-particle system
satisfies the following two conditions. If the relative distance between
any two particles belonging to the same group goes to infinity, the
probability density corresponding to the $N$-particle wave function
decays in the same way as a bound state. 
On the other hand, if the relative distance between any two particles
belonging to different groups tends towards infinity
(keeping the relative distances among 
all particles belonging to the same group unchanged), the probability
density remains finite similar to a scattering state.
If any wave function corresponding to a $N$-particle system
satisfies these two conditions, we define it as a clustered state.
 
Next, let us discuss how the conditions for constructing a clustered state
can be implemented for the case of the plane wave function (\ref{b7}). Since
any eigenvalue of the momentum operator ${\cal P}_N$ must be a real
quantity, Eq.~(\ref{b8}) is also obeyed for this case. To proceed further,
let us choose a specific value of $N$ given by $N=4$. 
For this case, the probability density (\ref{b9}) may be explicitly written as
\begin{equation} 
{|\rho_{1,2,\cdots ,4} (~ x_1,x_2, \cdots , x_4 ~)|}^2 ~=~ \exp~\{ 
2q_1y_1 + 2(q_1+q_2)y_2 + 2(q_1+q_2+q_3)y_3 \} ~. \label{b17} \end{equation}
Suppose, the conditions (\ref{b10}) for a bound state formation
are slightly modified for this case as 
\[ q_1<0, ~~~~q_1+q_2 =0, ~~~~q_1+q_2+q_3 <0 \, . \]
Taking into account this new condition, it is easy to see that when
$y_1 = x_2 - x_1$ or $y_3 = x_4 - x_3$ tends towards infinity, the probability
density in Eq.~(\ref{b17}) still decays like a bound state. 
On the other hand, when $y_2 = x_3-x_2$ tends towards
infinity, the probability density in Eq.~(\ref{b17}) remains finite. Hence,
the clusters of particles given by $\{1,2\}$ and $\{3,4\}$ satisfy all 
the criteria of a clustered state. Generalizing this specific example
in a straightforward way for any given values of $N$ and $\phi$,
we replace some of the inequalities in Eq.~(\ref{b10}) by equalities.
In this way, we find out the conditions for obtaining a clustered state from 
the plane wave function (\ref{b7}) as
\begin{eqnarray} 
&&~~~~~~~~~~~~~~~~~~~~~~~~\sum_{i=1}^l q_i = 0 \, , ~~{\rm
for}~ l~ \in {\Omega}_{N,\phi} \, 
,~~~~~~~~~~~~~~~~~~~~~~~~~~~~~~~~~~~~~~~~~~~~~~~~~~~
(2.18a) \nonumber \\
&& ~~~~~~~~~~~~~~~~~~~~~~~~\sum_{i=1}^l q_i ~<~ 0 \, , ~~ {\rm for}
~l~\in (\Omega_N - {\Omega}_{N,\phi})\, ,
~~~~~~~~~~~~~~~~~~~~~~~~~~~~~~~~~~~~~~~(2.18b)\nonumber \end{eqnarray}
\addtocounter{equation}{1}
where $\Omega_{N,\phi}$ denotes any non-empty proper subset of $\Omega_N$ and
$(\Omega_N - \Omega_{N,\phi})$ is the complementary set of $\Omega_{N,\phi}$. 

Next, we try to find the simplest
possible condition for which any Bethe state of the form
(\ref{b3}) would represent clusters of bound particles. 
Let us assume that the quasi-momenta corresponding to this Bethe state 
satisfy the relations (\ref{b12}). As a result, the coefficients of all plane
waves except (\ref{b7}) take the zero value within the Bethe state (\ref{b3}).
Thus, due to the relations (\ref{b12}), 
the Bethe state (\ref{b3}) would reduce to the plane wave function (\ref{b7}).
Next, we assume that the quasi-momenta corresponding to this 
Bethe state also satisfy the relations (\ref{b8}) and (2.18a,b).
Hence, the plane wave function (\ref{b7}) represents a clustered state.
Consequently, Eqs.~(\ref{b8}), (\ref{b12}) and (2.18a,b) together
yield a sufficient and simplest possible condition for which the Bethe state 
(\ref{b3}) would represent clusters of bound particles. Let us now analyse
this condition for the case of the derivative $\delta$-function Bose gas.
Using Eqs.~(\ref{b8}) and (\ref{b12}) along with the form of $A(k_l,k_m)$
given in (\ref{b5}) it is easy to see that, similar to the case of a 
localized bound state, the quasi-momenta associated with
clusters of bound particles can be written in the form (\ref{b13}) and the
imaginary parts of these quasi-momenta satisfy the relation (\ref{b15}).
With the help of Eq.~(\ref{b15}), we can recast Eqs.~(2.18a,b) as
\begin{eqnarray} 
&&~~~~~~~\chi ~\frac{\sin (l \phi)}{\sin \phi} ~\sin [(N-l) \phi] = 0
\, ,\quad {\rm
for}~~~ l~ \in {\Omega}_{N, \phi} \, ,~~~~~~~~~~~~~~~~~~~~~~~~~~~~~~~~~~~~~~~~
(2.19a) \nonumber \\
&&~~~~~~~\chi ~\frac{\sin (l \phi)}{\sin \phi} ~\sin [(N-l) \phi] ~>~ 0 \, ,
\quad {\rm for} ~~~l~\in (\Omega_N - \Omega_{N,\phi})\, .
~~~~~~~~~~~~~~~~~~~~~~~~~~~~(2.19b)\nonumber \end{eqnarray}
\addtocounter{equation}{1}
For any given values of $N$ and $\phi$,
the above equations clearly give the simplest
possible condition for which the Bethe state
(\ref{b3}) would represent clusters of bound particles. 
Let us assume that these equations are satisfied
for some values of $\phi$ and $N$, where $\Omega_{N, \phi}$ is given by
\begin{equation} 
\Omega_{N, \phi} = \{ l_1,l_2, \cdots , l_p \} \, , \label{b21} \end{equation}
with $1 \leq p < N-1$.  Then 
from Eq.~(\ref{b9}) it follows that 
the sets of particles given by
$\{ 1, \cdots ,l_1 \}, \{ l_1+1 , \cdots ,l_2 \}, \cdots \cdots ,
\{ l_{p-1}+1 , \cdots ,l_p \}, \{ l_{p}+1, \cdots , N \}$
represent $(p+1)$ number of clusters of bound particles. 
Moreover, the numbers of particles present within each of these clusters,
i.e., the size of the clusters, may be written in the form
\begin{equation} 
\{\!\{ \, l_1, l_2-l_1, \cdots , l_p-l_{p-1}, N- l_p \, \}\!\} \, .
\label{b22} \end{equation}

Since the quasi-momenta associated with both bound states and
clusters of bound particles are given by Eq.~(\ref{b13}), the momentum and
energy eigenvalues for clusters of bound particles can be derived in exactly
the same way as has been done earlier \cite{BBS03} for the case of a bound
state. Inserting the quasi-momenta given in Eq.~(\ref{b13}) to Eqs.~(2.6a,b),
we obtain the momentum eigenvalue as
\begin{equation} 
P ~=~ \hbar \chi ~\frac{\sin (N\phi)}{\sin \phi} ~, \label{b23} \end{equation}
and the energy eigenvalue as
\begin{equation} 
E ~=~ \frac{\hbar^2 \chi^2 \sin(2N \phi)}{\sin(2\phi)} ~. \label{b24}
\end{equation}

Let us make a comment at this place. 
Using Eq.~(\ref{b23}) we find that, for any given values of 
$N$ and $\phi$, the parameter $\chi $ is proportional to the total momentum 
$P$. Hence, due to Eqs.~(\ref{b9}) and (\ref{b15}) it follows that, 
in the limit of large relative distances between particle coordinates,
the probability density for a bound state wave function (or for a cluster 
within a clustered state) decays exponentially over a length which is 
inversely proportional to the total momentum $P$. This surprising result is a 
consequence of the fact that since we have set $2m=1$ and
the parameter $\eta$ is dimensionless in the Hamiltonian (\ref{a2}),
the only length scale appearing in an eigenstate of the Hamiltonian
is $\hbar /P$. Hence the decay length must be proportional to $\hbar/P$.

\section{Farey sequences and clusters of bound particles}
\renewcommand{\theequation}{3.{\arabic{equation}}}
\setcounter{equation}{0}
For the purpose of constructing clusters of bound particles
in the case of the derivative $\delta$-function Bose gas,
here our aim is to find out all possible solutions of Eqs.~(2.19a,b).
Some properties
of the Farey sequences \cite{ZZM00} in number theory will play a crucial
role in our analysis. Due to the existence of the parity transformation,
as mentioned in the earlier section, it is sufficient to concentrate on values
of $\phi$ lying in the range $0<\phi <\frac{\pi}{2}$. 
Within this range of $\phi$, $\sin\phi>0$ and hence the conditions (2.19a,b)
for forming clusters of bound particles reduce to 
\begin{eqnarray} 
&&~~~~~~~~~~\chi ~\sin (l \phi)~\sin [(N-l) \phi] = 0 \, , \quad \quad {\rm
~~for}~~~l~\in \Omega_{N, \phi} \, ,~~~~~~~~~~~~~~~~~~~~~~~~~~~~~
~~~~~~(3.1a) \nonumber \\
&&~~~~~~~~~~\chi ~\sin (l \phi)~\sin [(N-l) \phi] ~>~ 0 \, ,\quad\quad {\rm
for} ~~l ~ \in (\Omega_N - \Omega_{N, \phi})\,. 
\, ~~~~~~~~~~~~~~~~~~~~~~~~~~(3.1b)\nonumber \end{eqnarray}
\addtocounter{equation}{1}
It is easy to see that the condition (3.1a) would be
satisfied if and only if $\phi/\pi$ can be expressed in the form
\begin{eqnarray} \frac{\phi}{\pi} = \frac{a}{b}\,, \label{c2} \end{eqnarray}
where $\{a,b\}$ are relatively prime integers (i.e, the greatest common
divisor of $a$ and $b$ is 1), taking values within the ranges 
\begin{eqnarray} 
~~~~~~~~~~~~~~~~~~~~~~~~~~~~~~~~~~~~0< a < \frac{b}{2}\,, ~~~~2 < b \leq
N-1\,.
~~~~~~~~~~~~~~~~~~~~~~~~~~~~~~~~~~~~~~~(3.3a,b) \nonumber \end{eqnarray}
\addtocounter{equation}{1}
Due to Eq.~(3.3b) it is evident that, clusters of bound particles
can exist only for $N \geq 4$. In the following, we shall establish a
connection of the fractions $\phi/\pi$, given by Eqs.~(\ref{c2}) and
(3.3a,b), with the elements of Farey sequences in number theory. 

For a positive integer $N$, 
the Farey sequence is defined to be the set of all
the fractions $a/b$ in increasing order such that (i) $0 \le a \le b \le N$,
and (ii) \{$a, b$\} are relatively prime integers \cite{ZZM00}. 
The Farey sequences for the first few values of $N$ are given by
\begin{eqnarray} 
F_1: & & \quad \frac{0}{1} ~~~~\frac{1}{1} \nonumber \\
F_2: & & \quad \frac{0}{1} ~~~~\frac{1}{2} ~~~~\frac{1}{1} \nonumber \\
F_3: & & \quad \frac{0}{1} ~~~~\frac{1}{3} ~~~~\frac{1}{2} ~~~~\frac{2}{3}
~~~~
\frac{1}{1} \nonumber \\
F_4: & & \quad \frac{0}{1} ~~~~\frac{1}{4} ~~~~\frac{1}{3} ~~~~\frac{1}{2}
~~~~
\frac{2}{3} ~~~~\frac{3}{4} ~~~~\frac{1}{1} \nonumber \\
F_5: & & \quad \frac{0}{1} ~~~~\frac{1}{5} ~~~~\frac{1}{4} ~~~~\frac{1}{3}
~~~~
\frac{2}{5} ~~~~\frac{1}{2} ~~~~\frac{3}{5} ~~~~\frac{2}{3} ~~~~
\frac{3}{4} ~~~~\frac{4}{5} ~~~~\frac{1}{1}
\label{c4} 
\end{eqnarray}
These sequences enjoy several properties, of which we list the relevant ones
below. \\
\noindent (i) Let $a/b, a'/b'$ are two fractions appearing in the Farey
sequence $ F_N$. Then $a/b < a'/b' ~(~ a'/b' < a/b~) $ are two successive
fractions in $F_N$, if and only if the following two conditions are
satisfied:
\begin{eqnarray} 
&&~~~~~~~~~~~~~~~~~~~~~~~~~~~~~~~~~~~~~~~a' b ~-~ a b' ~=~ 1~ (-1)~, ~~~~
~~~~~~~~~~~~~~~~~~~~~~~~~~~~~~~~~~~~~~(3.5a) \nonumber \\
&&~~~~~~~~~~~~~~~~~~~~~~~~~~~~~~~~~~~~~~~b+b'~>~N\,.
 ~~~~~~~~~~~~~~~~~~~~~~~~~~~~~~~~~~~~~~~~~~~~~~~~~~~~~~~
(3.5b) \nonumber 
\end{eqnarray}
\addtocounter{equation}{1}
It then follows that both $a$ and $b'$ are relatively prime to $a'$ and $b$.

\noindent (ii) For $N \ge 2$, if $n/N$ is a fraction appearing somewhere in the
sequence $F_N$ (this implies that $N$ and $n$ are relatively prime according
to
the definition of $F_N$), then the fractions $a_1/b_1$ and $a_2/b_2$
appearing 
immediately to the left and to the right respectively of $n/N$ satisfy
\begin{eqnarray} 
a_1 ~,~ a_2 ~\le ~n ~, \quad {\rm and} \quad a_1 ~+~ a_2 ~=~ n ~, \nonumber
\\
b_1 ~,~ b_2 ~<~ N ~, \quad {\rm and} \quad b_1 ~+~ b_2 ~=~ N ~. \label{c6}
\end{eqnarray}

To apply the above mentioned Farey sequence in the present context,
let us define a subset of $F_N$ as 
\begin{eqnarray} 
{F}^\prime_N = \left \{ \left.\frac{a}{b} ~\right| ~~\frac{a}{b} \in
F_N,
 ~~~\frac{0}{1}<\frac{a}{b} < \frac{1}{2} \right\}\,, \label{c7} 
\end{eqnarray}
and a subset of $F^\prime_N$ as 
\begin{eqnarray} 
F''_N = \left\{ \left.\frac{n}{N} ~\right| ~~ \frac{n}{N} \in
F^\prime_N
\right \}\,. \label{c8} \end{eqnarray}
Using these definitions of various subsets of a Farey sequence,
it is easy to show that 
\begin{eqnarray} 
F^\prime_N = F^\prime_{N-1} \cup F''_N\,. \label{c9} \end{eqnarray}
Furthermore, it is worth noting that,
 Eqs.~(\ref{c2}) and (3.3a,b) can equivalently be expressed as
\begin{eqnarray} \frac{\phi}{\pi} \in F^\prime_{N-1} \,. \label{c10} 
\end{eqnarray} 
Consequently, it follows that the condition (3.1a) for cluster
formation is obeyed if and only if $\phi/\pi \in F'_{N-1}$.
In this context it may be observed that, due to Eq.~(\ref{c9}),
all the elements of $F^\prime_{N-1}$ are also present in $F^\prime_N$.
By using such an embedding of $F^\prime_{N-1}$ into $F^\prime_N$,
we find that any fraction $a/b \in F'_{N-1}$ belongs to one of the
four distinct classes, which are defined in the following:
\vskip .11 cm 
\noindent
I. At least one of the fractions nearest to $a/b$ (from either the left or
the right side) in the sequence $F^\prime_N$ lies in the set $F''_N$.
Then, from a property of the Farey sequences, it follows that
$\{b, N\}$ are relatively prime integers in this case.
\\
II. None of the nearest fractions of $a/b$ (from the left or right
side) in the sequence $F'_N$ lies in the 
set $F''_N$, and $\{b, N\}$ are relatively prime integers.
\\
III. $N$ is divisible by $b$. Clearly, $\{b, N\}$ are not relatively prime
integers in this case.
\\
IV. $N$ is not divisible by $b$, and $\{b,N\}$ are not relatively prime
integers.
\vskip .11 cm
\noindent
To demonstrate the above mentioned classification through an
example, let us choose $N=6$. For this case, the sets
$F^{\prime}_5$, $F^{\prime}_6$ and $F''_6$ are given by
\begin{eqnarray}
F^\prime_5 :& &\frac{1}{5}~~~~\frac{1}{4}~~~~\frac{1}{3}~~~~
\frac{2}{5} \nonumber \\
F^\prime_6 :& &\frac{1}{6}~~~~\frac{1}{5}~~~~\frac{1}{4}~~~~~
\frac{1}{3}~~~~\frac{2}{5}\,; ~~~~~F''_6 :~~ \frac{1}{6}\,.\nonumber 
\end{eqnarray}
Using the embedding of $F^{\prime}_5$ into $F^{\prime}_6$,
it is easy to verify that each fraction in $F^\prime_5$ falls under one
of the four classes discussed above. More precisely, the fractions 1/5,
1/4, 1/3 and 2/5 belong to type I, type IV, type III and type II
respectively.
Returning back to the general case we note that, for any fraction $a/b \in
F'_{N-1} \,$, $\{b,N\}$ are either relatively prime integers 
or not relatively prime integers. 
If $\{b,N\}$ are relatively prime integers, then it is obvious that
$a/b$ must be an element of either type I or type II. On the other hand,
if $\{b,N\}$ are not relatively prime integers, then 
$a/b$ must be an element of either type III or type IV. In this way,
one can show that any fraction $a/b \in F'_{N-1}$ belongs to one of these
four distinct classes. 

Through a lengthy analysis which 
uses the properties (3.5a,b) and (\ref{c6}) 
of a Farey sequence, we have shown that fractions
of types I and III belonging to $F'_{N-1}$ satisfy the relation
(3.1b), while type II and type IV fractions do not satisfy (3.1b)
\cite{BBS13}. Hence, 
clusters of bound particles are formed for the case of derivative 
$\delta$-function Bose gas,
if and only if the corresponding coupling constant
$\phi/\pi = a/b \in F'_{N-1}$ is a fraction of type I or type III.

\section{Some properties of clusters of bound particles}
\renewcommand{\theequation}{4.{\arabic{equation}}}
\setcounter{equation}{0}

In this section, we shall compute the 
number of clusters present within a Bethe state representing clusters of
bound particles and the sizes of these
clusters (i.e., number of bound particles present in each of these
clusters). Subsequently, we shall analyse the behavior of these 
clusters of bound particles under small variations of the coupling constant.

\noindent \subsection{Sizes of the clusters of bound particles}

Let us first consider clusters of bound particles
when $\phi/\pi = a/b$ is taken as any fraction of type I within the set
$F'_{N-1}$. In section 2 we have seen that, 
to find the number and sizes of the clusters within a Bethe state, 
we have to determine the set $\Omega_{N,\phi}$ for which Eq.~(3.1a)
is satisfied. Since $\{N,b\}$ are relatively prime integers
for any fraction of type I, we can express $N$ as
\begin{eqnarray} N=pb+r \, , \label{d1} \end{eqnarray}
where $1 \leq r \leq b-1 $. Hence, for the discrete variable $l$ taking values
within the set $\Omega_N$, the zero points of the functions $\sin l\phi$ and
$\sin(N-l)\phi$ are respectively given by the sets
\begin{eqnarray}
~~~~~~~~~~~~~~~~~~~~~S_1 \equiv \{b, 2b, \cdots \cdots , pb\}, ~~~ S_2 
\equiv \{ N-b, N-2b,
\cdots \cdots , N - pb \}\, . ~~~~~~~~~~~(4.2a,b) \nonumber \end{eqnarray}
\addtocounter{equation}{1}
\hskip -.18 cm
Combining the sets $S_1$ and $S_2$ by using Eqs.~(\ref{d1}) and (4.2a,b),
we obtain $\Omega_{N,\phi}$ as
\begin{eqnarray} \Omega_{N, \phi} = S_1\cup S_2 = \{r,~b,~r+b,~2b, \cdots
\cdots,~r+(p-1)b,~pb\}\,. \label{d3} \end{eqnarray}
Comparing (\ref{d3}) with (\ref{b21}), and also using (\ref{b22}),
it is easy to see that the sizes of the clusters are given by
\begin{eqnarray} 
\{ \!\{r,~b-r,~r,~b-r,~\cdots \cdots,~r,~b-r,~r\}\!\}\,. \label{d4}
\end{eqnarray}
Hence, for any fraction of type I, the corresponding Bethe state contains
$(p+1)$ number of clusters of size $r$ and $p$ number of clusters of size
$(b-r)$. 
Next, by using the method of contradiction, we would like to show that these
two possible sizes of the clusters, i.e., $r$ and $(b-r)$, must be
relatively prime integers. To this end, let us first assume that $b$ and $r$
are not relatively prime integers. Therefore, we can write $b$ and $r$ as
$b=\alpha b'$ and $r=\alpha r'$, where $\alpha > 1$. Substituting these values
of $b$ and $r$ in Eq.~(\ref{d1}), we find that \[ N = \alpha (pb' + r')\, .  \]
Thus $\alpha$ is a common factor of $N$ and $b$. However, this result 
contradicts the fact that $\{N,b\}$ must be relatively prime integers for any
fraction of type I. Hence it is established that $\{b,r\}$ are relatively
prime integers. From this relation, it trivially follows that $\{r, b-r \}$
are relatively prime integers and, in particular, $r \neq b-r$.
Consequently from Eq.~(\ref{d4}) we find that, for any fraction of type I,
the corresponding Bethe state contains heterogeneous clusters of two 
different sizes. As a special case, let us consider any fraction of type I
with denominator satisfying the relation $b > N/2$. Due to Eq.~(\ref{d1}) it
follows that, $p=1$, $r=N-b$ and $b-r = 2b - N$ for this case. 
Hence, the corresponding Bethe state contains 
two clusters of the size $(N-b)$ and one cluster of the size $(2b - N)$.

Next, we consider the clusters of bound particles corresponding to any
fraction of type III. In this case $N$ can be written as $N=pb$, 
where $p$ is an integer greater than one. Consequently, for the variable $l$
taking value within the set $\Omega_N$, the zero points of the functions
$\sin l\phi$ and $\sin(N-l)\phi$ coincide with each other and yield 
$\Omega_{N, \phi}$ as 
\begin{eqnarray} 
\Omega_{N, \phi}= \{ b,~2b,~3b,\cdots \cdots ,(p-1)b \}\,. \label{d5}
\end{eqnarray}
Hence $p$ number of clusters are formed in this case.
Comparing (\ref{d5}) with (\ref{b21}), and also using (\ref{b22}),
it is easy to see that each cluster of this type has the size $b$.
In other words, the corresponding Bethe state (\ref{b3}) contains $N/b$
of homogeneous clusters, each of which is made of $b$ number of bound 
particles.
In Table 1 we show all the fractional values of $\phi /\pi$ for which clusters
of bound particles exist within the range of $N$ given by $4 \leq N \leq
10$, the types of these fractions 
and the sizes of the corresponding clusters using the notation of
Eq.~(\ref{b22}).

\vspace{1.1cm}
\begin{table}[htb]
\begin{center}
\begin{tabular}
{|c|c|c|c|} \hline
$N$ & Value of $\phi/\pi$ & Type 
& Size of the clusters \\
\hline
4 &$1/3$ & I 
& $\{ \! \{ 1,2,1 \} \! \}$\\
5 & $1/4$ & I 
& $\{ \!\{1,3,1\}\!\}$\\
5 & $1/3$ & I 
& $\{\!\{2,1,2\}\!\}$\\
6 & $1/5$ & I 
& $\{\!\{1,4,1 \}\!\}$\\
6 & $1/3$ & III 
& $\{\!\{3,3\}\!\}$ \\
7 & $1/6$ & I 
& $\{\!\{1, 5, 1 \}\!\}$\\
7 & $1/4$ & I 
& $\{\!\{3,1,3 \}\!\}$\\
7 & $1/3$ & I 
& $\{\!\{1,2,1,2,1 \}\!\}$ \\
7 & $2/5$ & I 
& $\{\!\{2,3,2 \}\!\}$\\
8 & $1/7$ & I 
& $\{\!\{1,6,1\}\!\}$\\
8 & $1/4$ & III 
& $\{\!\{ 4, 4 \}\!\}$ \\
8 & $1/3$ & I 
& $\{\!\{ 2, 1, 2, 1, 2 \}\!\}$\\
8 & $2/5$ & I 
& $\{\!\{3, 2, 3 \}\!\}$ \\
9 & $1/8$ & I 
& $\{\!\{1, 7, 1 \}\!\}$\\
9 & $1/5$ & I 
& $\{\!\{ 4, 1, 4 \}\!\}$\\
9 & $1/4$ & I 
& $\{\!\{ 1, 3, 1, 3, 1 \}\!\}$ \\
9 & $1/3$ & III 
& $\{\!\{3, 3, 3\}\!\}$ \\
9 & $3/7$ & I 
& $\{\!\{ 2, 5, 2 \}\!\}$\\
10 & $1/9$ & I 
& $\{\!\{ 1,8,1 \}\!\}$ \\
10 & $1/5$ & III 
& $\{ \! \{ 5, 5 \}\!\}$ \\
10 & $2/7$ & I 
& $\{\!\{ 3,4,3 \}\!\}$ \\
10 & $1/3$ & I 
& $\{\!\{ 1, 2, 1, 2, 1, 2, 1 \}\!\}$ \\
10 & $2/5$ & III 
& $\{\!\{ 5, 5 \}\!\}$ \\ 
\hline
\end{tabular}
\end{center} 
\caption{\label{1}The fractional values of $\phi /\pi$ for which clusters of 
bound particles exist for $4 \leq N \leq 10$, the types of these fractions 
and the sizes of the corresponding clusters are shown.}
\vskip .2 cm 
\end{table}


\noindent
\subsection{Stability of clusters of bound particles and binding
energies}


Here, we shall discuss about the stability of clusters of bound particles 
under infinitesimal variations of the coupling constant. For 
$N \geq 4$, let us choose any specific 
value of the coupling constant $\phi$ within the range $0<\phi<\frac{\pi}{2}$
such that clusters of bound particles are formed.
If one increases or decreases this value of $\phi$ by an infinitesimal
amount, it is obvious that all inequalities in Eq.~(3.1b) would
continue to be satisfied and 
all equalities in Eq.~(3.1a) would be transformed into some inequalities. 
Consequently, clusters of bound particles cease to exist even
for a very small change of the coupling constant. One of the following 
two different cases can occur in such a situation. In the first case,
at least one of the equalities in Eq.~(3.1a) is transformed 
into an inequality of the form 
\begin{equation} \chi ~\sin (l \phi)~\sin [(N-l) \phi] ~<~ 0 \, . \label{d6} 
\end{equation}
It is evident that, the probability density of the corresponding
Bethe state (\ref{b3}) would diverge if the relative distance between
at least one pair of particle coordinates tends towards infinity.
As a result, this Bethe state becomes ill-defined and disappears from the
Hilbert space of the Hamiltonian (\ref{a2}) of derivative the $\delta$-function
Bose gas. So we may say that clusters of bound particles become unstable
in this case. Let us now consider the second case, for which all of
equalities in Eq.~(3.1a) are transformed into inequalities of the form 
\begin{equation} 
\chi ~\sin (l \phi)~\sin [(N-l) \phi] ~>~ 0\, , \label{d7} \end{equation}
due to an infinitesimal change of the coupling constant $\phi$. It is evident
that, for this case, Eqs.~(3.1a,b) are transformed to Eq.~(\ref{b16}) within 
the range of $\phi$ given by $0<\phi<\frac{\pi}{2}$.
As a result, clusters of bound particles merge with each other and
produce a localized bound state containing only one cluster of particles.
Therefore, in this second case, we may say that clusters of bound particles
turn into a localized bound state with only one cluster. 

Using the above mentioned procedure,
we have found that \cite{BBS13} clusters of bound particles associated with
fractions of type III become unstable and cease to exist 
for any small change of the coupling constant. 
On the other hand, for the case of fractions of type I, clusters of 
bound particles transmute to a localized bound state containing only one
cluster if the value of $\phi$ is slightly changed towards the direction of 
the nearest fraction $n/N$. However, such clusters of bound particles 
become unstable if the value of $\phi$ is slightly changed 
towards the opposite direction.
Consequently, fractions of type I lie at the end points of the bands 
containing localized bound states.

\section{Conclusion}
In this article, we have explored how
clusters of bound particles can be constructed in the simplest possible
way for the case of an exactly solvable derivative $\delta$-function Bose gas.
To this end, we consider a sufficient condition for which Bethe states of
the form in (\ref{b3}) would lead to clusters of bound particles. It is found 
that this sufficient condition can be satisfied by taking the quasi-momenta 
of the corresponding Bethe state to be equidistant points on a {\it single}
circle having its centre at the origin of the complex momentum plane.
Furthermore, the coupling constant ($\phi$) and the total number of particles 
($N$) of this derivative $\delta$-function Bose gas must satisfy the relations
in (3.1a,b). For any given $N\geq 4$, it is found that Eq.~(3.1a) is satisfied
if $\phi/\pi$ takes any value within the set $F'_{N-1}$, which is 
a subset of the Farey sequence $F_{N-1}$. Then we classify all fractions
belonging to the set $F'_{N-1}$ into four types. It turns out that fractions
of types I and III belonging to $F'_{N-1}$ satisfy the remaining relation
(3.1b), while type II and type IV fractions do not satisfy (3.1b).
Consequently, clusters of bound particles can be constructed for the
derivative $\delta$-function Bose gas only for special values of $\phi/\pi$
given by the fractions of types I and III within the set $F'_{N-1}$.

We have also computed the sizes of the above mentioned clusters of bound
particles, i.e., the number of particles present within each of these 
clusters. It is found that any fraction of type I within the set $F'_{N-1}$ 
leads to heterogeneous clusters of bound particles having two different sizes. 
On the other hand, any fraction of type III within the set $F'_{N-1}$
leads to homogeneous clusters of bound particles having only one size.
Interestingly, clusters of bound particles associated with fractions of
type I and type III transform in rather different ways under a small
variation of the coupling constant. For example, it is found that
clusters of bound particles associated with fractions of type III
cease to exist for any small change of the coupling constant. On the other
hand, clusters of bound particles corresponding to fractions of type I turn
into localized bound states consisting of a single cluster if the value of 
the coupling constant is slightly increased or decreased. 

In this paper we have analyzed a particular type of sufficient condition for 
constructing clusters of bound particles in the case of the derivative 
$\delta$-function Bose gas in the finite volume limit. However, in future,
it might be interesting to explore other possible ways of constructing 
clusters of bound particles for this exactly solvable system.
As is well known, many properties of the complex quasi-momenta for the 
case of the $\delta$-function Bose gas can be derived relatively easily 
by using the solutions of the corresponding finite volume Bethe ansatz 
equations. A similar analysis of the finite volume Bethe ansatz equations
may be helpful in finding all possible clusters of bound particles 
for the derivative $\delta$-function Bose gas in the infinite volume limit. 

\vspace{1.1cm}
\noi {\bf Acknowledgments}
\vspace{.2cm} 

B.B.M. thanks the organizers of `The XXIInd International Conference on 
Integrable Systems and quantum symmetries (ISQS-22)' for inviting to present
this work and J.S. Caux for fruitful discussions. B.B.M. also thanks the Abdus 
Salam International Centre for Theoretical Physics for a Senior Associateship, 
which partially supported this work. D.S. thanks the Department of Science and 
Technology, India for financial support through the grant SR/S2/JCB-44/2010.



\end{document}